\documentclass[twocolumn,showpacs,aps,prl,superscriptaddress]{revtex4}

\usepackage{graphicx}
\usepackage{dcolumn}
\usepackage{amsmath}
\usepackage{epsfig}

\RequirePackage{xspace}

\usepackage{relsize}
\def\babar{\mbox{\slshape B\kern-0.1em{\smaller A}\kern-0.1em
    B\kern-0.1em{\smaller A\kern-0.2em R}}}

\def\Kbar  {\kern 0.2em\overline{\kern -0.2em K}{}\xspace}

\def\Kz    {\ensuremath{K^0}\xspace}
\def\Kzb   {\ensuremath{\Kbar^0}\xspace}
\def\KzKzb {\ensuremath{\Kz \kern -0.16em \Kzb}\xspace}
\def\Kp    {\ensuremath{K^+}\xspace}
\def\Km    {\ensuremath{K^-}\xspace}

\def\KpKm  {\ensuremath{\Kp \kern -0.16em \Km}\xspace}
\def\KS    {\ensuremath{K^0_{\scriptscriptstyle S}}\xspace}

\def\Dbar    {\kern 0.2em\overline{\kern -0.2em D}{}\xspace}

\def\Dz      {\ensuremath{D^0}\xspace}
\def\Dzb     {\ensuremath{\Dbar^0}\xspace}
\def\DzDzb   {\ensuremath{\Dz {\kern -0.16em \Dzb}}\xspace}
\def\Dp      {\ensuremath{D^+}\xspace}
\def\Dm      {\ensuremath{D^-}\xspace}

\def\DpDm    {\ensuremath{\Dp {\kern -0.16em \Dm}}\xspace}

\def\Bbar    {\kern 0.18em\overline{\kern -0.18em B}{}\xspace}

\def\Bz      {\ensuremath{B^0}\xspace}
\def\Bzb     {\ensuremath{\Bbar^0}\xspace}
\def\BzBzb   {\ensuremath{\Bz {\kern -0.16em \Bzb}}\xspace}
\def\Bu      {\ensuremath{B^+}\xspace}
\def\Bub     {\ensuremath{B^-}\xspace}

\def\BpBm    {\ensuremath{\Bu {\kern -0.16em \Bub}}\xspace}

\def\BorBbar    {\kern 0.18em\optbar{\kern -0.18em B}{}\xspace}
\def\DorDbar    {\kern 0.18em\optbar{\kern -0.18em D}{}\xspace}
\def\KorKbar    {\kern 0.18em\optbar{\kern -0.18em K}{}\xspace}

\def\jpsi     {\ensuremath{{J\mskip -3mu/\mskip -2mu\psi\mskip 2mu}}\xspace}

\def\etac     {\ensuremath{\eta_c}\xspace}

\mathchardef\Upsilon="7107
\def\Y#1S{\ensuremath{\Upsilon{(#1S)}}\xspace}

\mathchardef\Deltares="7101
\mathchardef\Xi="7104
\mathchardef\Lambda="7103
\mathchardef\Sigma="7106
\mathchardef\Omega="710A

\def\Deltabar{\kern 0.25em\overline{\kern -0.25em \Deltares}{}\xspace}
\def\Lbar{\kern 0.2em\overline{\kern -0.2em\Lambda\kern 0.05em}\kern-0.05em{}\xspace}
\def\Sigbar{\kern 0.2em\overline{\kern -0.2em \Sigma}{}\xspace}
\def\Xibar{\kern 0.2em\overline{\kern -0.2em \Xi}{}\xspace}
\def\Obar{\kern 0.2em\overline{\kern -0.2em \Omega}{}\xspace}
\def\Nbar{\kern 0.2em\overline{\kern -0.2em N}{}\xspace}
\def\Xb{\kern 0.2em\overline{\kern -0.2em X}{}\xspace}

\newcommand{\tev}{\ensuremath{\mathrm{\,Te\kern -0.1em V}}\xspace}
\newcommand{\gev}{\ensuremath{\mathrm{\,Ge\kern -0.1em V}}\xspace}
\newcommand{\mev}{\ensuremath{\mathrm{\,Me\kern -0.1em V}}\xspace}
\newcommand{\kev}{\ensuremath{\mathrm{\,ke\kern -0.1em V}}\xspace}
\newcommand{\ev}{\ensuremath{\mathrm{\,e\kern -0.1em V}}\xspace}
\newcommand{\gevc}{\ensuremath{{\mathrm{\,Ge\kern -0.1em V\!/}c}}\xspace}
\newcommand{\mevc}{\ensuremath{{\mathrm{\,Me\kern -0.1em V\!/}c}}\xspace}
\newcommand{\gevcc}{\ensuremath{{\mathrm{\,Ge\kern -0.1em V\!/}c^2}}\xspace}
\newcommand{\mevcc}{\ensuremath{{\mathrm{\,Me\kern -0.1em V\!/}c^2}}\xspace}

\def\mus  {\ensuremath{\rm \,\mus}\xspace}

\def\mus        {\ensuremath{\,\mu{\rm s}}\xspace}    

\def\to                 {\ensuremath{\rightarrow}\xspace}

\def\pep2{PEP-II}

\def\gsim{{~\raise.15em\hbox{$>$}\kern-.85em
          \lower.35em\hbox{$\sim$}~}\xspace}
\def\lsim{{~\raise.15em\hbox{$<$}\kern-.85em
          \lower.35em\hbox{$\sim$}~}\xspace}

\xspace

\newcommand{\jprlBase}       {Phys.\ Rev.\ Lett.\xspace}
\newcommand{\jprBase}        {Phys.\ Rev.\xspace}
\newcommand{\jplBase}        {Phys.\ Lett.\xspace}
\newcommand{\nimBaseA}       {Nucl.\ Instr.\ Meth.\xspace}

\newcommand{\nima}      [1]  {\nimBaseA~A~{\bf #1}}

\newcommand{\plb}       [1]  {\jplBase\ B~{\bf #1}}
\newcommand{\prep}      [1]  {{Phys.\ Rep.\ {\bf #1}}}
\newcommand{\jprl}      [1]  {\jprlBase\ {\bf #1}}
\newcommand{\jprd}      [1]  {\jprBase\ D~{\bf #1}}

\def\jetset74   {\mbox{\tt Jetset \hspace{-0.5em}7.\hspace{-0.2em}4}\xspace}

\newcommand{\BABARPubYear}    {03}
\newcommand{\BABARPubNumber}  {023}

\newcommand{\SLACPubNumber} {10245}

\def\figurebox#1#2#3{%
    \def\arg{#3}%
    \ifx\arg\empty
    {\hfill\vbox{\hsize#2\hrule\hbox to #2{\vrule\hfill\vbox to #1{\hsize#2\vfill}\vrule}\hrule}\hfill}%
    \else
    {\hfill\epsfbox{#3}\hfill}%
    \fi}
\def\kskpi {\ensuremath{\KS K^{\pm}\pi^{\mp}}\xspace}
\begin{document}

\preprint{\babar-PUB-\BABARPubYear/\BABARPubNumber} 
\preprint{SLAC-PUB-\SLACPubNumber} 

\begin{flushleft}
\babar-PUB-\BABARPubYear/\BABARPubNumber\\
SLAC-PUB-\SLACPubNumber\\
\end{flushleft}

\title{
{\large \bf
Measurements of the Mass and Width of the \etac Meson and 
of an $\etac(2S)$ Candidate} 
}

\author{B.~Aubert}
\author{R.~Barate}
\author{D.~Boutigny}
\author{J.-M.~Gaillard}
\author{A.~Hicheur}
\author{Y.~Karyotakis}
\author{J.~P.~Lees}
\author{P.~Robbe}
\author{V.~Tisserand}
\author{A.~Zghiche}
\affiliation{Laboratoire de Physique des Particules, F-74941 Annecy-le-Vieux, France }
% ...
\collaboration{The \babar\ Collaboration}
\noaffiliation

\date{\today}

\begin{abstract}
The mass $m_{\etac}$ and total width $\Gamma_{tot}^{\etac}$ of the \etac 
meson have been measured in two-photon interactions at the SLAC $e^+ e^-$ 
asymmetric $B$-Factory with the \babar\ detector. With a sample of approximately 
2500 reconstructed $\etac \to \kskpi$ decays in 88 fb$^{-1}$ of data, the 
results are $m_{\etac}$ = 2982.5 $\pm$ 1.1 (stat) $\pm$ 0.9 (syst)\mevcc and
$\Gamma_{tot}^{\etac}$ = 34.3 $\pm$ 2.3 (stat) $\pm$ 0.9 (syst)\mevcc.
Using the same decay mode, a second resonance with $112\pm24$  events
is observed with a mass of 3630.8 $\pm$ 3.4 (stat) $\pm$ 1.0 (syst)\mevcc
 and width of 17.0 $\pm$ 8.3 (stat) $\pm$ 2.5 (syst)\mevcc.
This observation is consistent with expectations for the $\etac(2S)$ state.
\end{abstract}


\maketitle
%
%

The mass and width of the \etac meson ($J^{PC}=0^{-+}$), 
the lowest lying state of charmonium, are not as well established as those of
the \jpsi meson. The world average~\cite{pdg} of the total width is
$\Gamma_{tot}^{\etac} = 16.0^{+3.6}_{-3.2}$\mevcc, with individual 
measurements ranging from 7\mevcc to 27\mevcc with large errors. Recent 
measurements~\cite{etac} extend from 17\mevcc to 29\mevcc.

A radial excitation of the \etac, the $\etac(2S)$ state, is predicted 
by heavy quark potential models to lie below the $D \bar{D}$ threshold 
\cite{theor}. The hyperfine separations ($\etac,\jpsi$) and  
($\etac(2S),\psi(2S)$) are directly related to the spin-spin interaction. 
These calculations predict the mass splitting
$m_{\psi(2S)}-m_{\etac(2S)}$ to be in the range 42--103\mevcc.
The Crystal Ball Collaboration~\cite{edw82} observed a peak at 91 $\pm$ 5\mev, 
in the inclusive photon spectrum of $\psi(2S)$ decays, 
with a width $\Gamma \leq 8$\mev (95\% confidence level). 
This peak was considered most likely to be due to $\psi(2S) \to
\etac(2S) \gamma$, with the  $\etac(2S)$ state having a mass of 3594 
$\pm$ 5\mevcc. The Belle Collaboration recently reported signals attributed 
to the $\etac(2S)$ state, but with substantially higher masses: 
for the $\KS K^- \pi^+$ mass distribution in exclusive 
$B \to K \KS K^- \pi^+$ decays~\cite{cho02}, they measured 
3654 $\pm$ 6 (stat) $\pm$ 8 (syst)\mevcc and $\Gamma \leq 55$\mevcc
(90\% confidence level); from a signal observed in the inclusive 
\jpsi spectrum in $e^+ e^-$ annihilation~\cite{abe02}, 
they measured  3622 $\pm$ 12\mevcc.
This state was unsuccessfully searched for in $p \bar{p} \to X 
\to \gamma \gamma$~\cite{fermi} and $\gamma \gamma \to hadrons$ 
\cite{cern}. However, an estimate~\cite{bar96} of the two-photon 
production rate of the $\etac(2S)$ suggested that this meson could be
identified in the current $e^+ e^-$ $B$-factories.

In this analysis we measure the masses and widths of the \etac and of a state 
interpreted as the $\etac(2S)$ meson,  by reconstructing 
$\gamma \gamma \to X \to \kskpi$ ($\KS \to \pi^+ \pi^-$) events in the 
 \babar\  detector at the PEP-II energy-asymmetric $e^+ e^-$ storage ring
at SLAC. The data sample was collected both on and slightly below the 
$\Upsilon(4S)$ resonance, and corresponds to an integrated luminosity 
of 88 fb$^{-1}$. 

The \babar\  detector is described in detail in reference~\cite{bab02}. The 
momenta of charged particles are measured and their trajectories reconstructed
with two detector systems located in a 1.5 T solenoidal magnetic field: 
a five-layer, double-sided silicon strip vertex tracker and a 40-layer drift 
chamber. Both devices provide $dE/dx$ measurement. Charged particle 
identification  is provided by a detector of internally reflected 
Cherenkov light,  complemented by the $dE/dx$ measurement.
The energies of electrons and photons are measured in a calorimeter consisting 
of 6580 CsI(Tl) crystals. 

The mesons are formed by the interaction of two virtual photons. Since the 
$e^+$ and $e^-$ scatter through too small an angle to be
 detected, the two photons are  
quasi-real and nearly aligned with the incident beams. 
A preselected sample comprises events having four charged tracks with 
a net zero charge and with total laboratory energy less than 9\gev. This 
removes most events coming from $B$ meson decays. 

A further selection of events is aimed at maximizing the ratio 
$S/\sqrt{(S+B)}$, where $S$ is the signal and $B$ the background, both taken 
within a $\pm 50\mevcc$ window around the \etac peak. Events with total 
transverse momentum in the center-of-mass  greater than 1.05\gevc or  with 
total energy of neutral particles  greater than  0.7\gev are rejected.
In order to identify $\etac \to \kskpi$ events, decays with one 
$\KS \to \pi^+ \pi^-$ candidate that lies within the  window 
$0.482 \leq M(\KS) \leq 0.512$\gevcc are selected. Of the two remaining tracks, 
we require that one and only one be identified as a kaon; the other one is 
assumed to be a pion. 
The angle between the \KS momentum and its flight path, as determined by
the \KS and $K^{\pm} \pi^{\mp}$ vertices, is required to be small
($\cos{\theta(\KS)} \geq 0.992$). Finally the \kskpi vertex is fitted, with the 
\KS  mass constrained to the world average value~\cite{pdg}. 

\begin{figure}[!htb]
\centerline{ \psfig{file=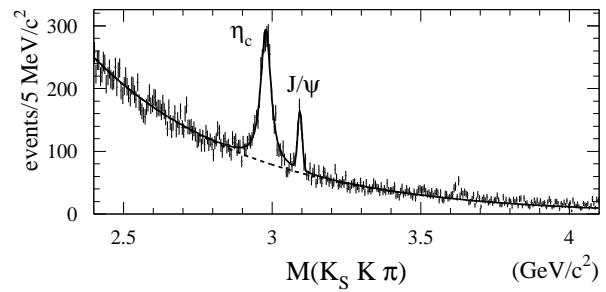,width=8.cm} }
\caption{ The (\kskpi) mass spectrum fitted (solid
line) to \etac + \jpsi + background, as explained in the text. 
The dashed line shows the background component of this fit.}
\label{fit}
\end{figure}

The resulting \kskpi mass spectrum is shown in Fig.~\ref{fit}, with a
large peak at the \etac mass and a smaller peak at the \jpsi mass. Although 
the \jpsi cannot be produced in two-photon fusion, it is expected to 
be produced with hard photon emission by initial state radiation (ISR).
The boost of the asymmetric collider brings  the decay products of 
\jpsi mesons travelling in the backward direction into the acceptance of the 
detector. 

A thorough understanding of the experimental resolution is essential to 
determine the width of the \etac meson. The resolution for the \jpsi can be 
inferred from  data since its natural width is negligible. This is not the 
case for the \etac, which has a natural width somewhat larger than the detector 
resolution. To help determine the resolution for the \etac, Monte Carlo 
calculations were performed. The generator~\cite{paa} used to simulate 
 $\gamma \gamma \to \etac \to \kskpi$ events 
applies the formalism of Budnev {\em et al.}~\cite{bud75}
to calculate the cross-section for the process $e^+ e^- \to e^+ e^- 
\gamma \gamma \to e^+ e^- \etac$. Monte Carlo calculations were also performed 
to generate  \jpsi events produced in $e^+ e^-$ annihilation with initial 
state radiation. Both \etac and  \jpsi were assumed to decay into \kskpi with 
a phase-space distribution. In the Monte Carlo simulation, the reconstructed 
\etac and \jpsi masses are both shifted by $-1.1\mevcc$ (with statistical 
errors of 0.1\mevcc and 0.2\mevcc respectively)
from their generated values. This bias does 
not affect the mass difference $m_{\jpsi}-m_{\etac}$. The mass resolution is 
estimated by fitting the distribution of the difference between reconstructed 
mass and generated mass to a Gaussian function. Its standard deviation is found 
to be 7.3 $\pm$ 0.1\mevcc for the \etac and 8.1 $\pm$ 0.2\mevcc for the \jpsi. 

To determine the mass and width of the \etac, an unbinned maximum likelihood 
fit to the \kskpi mass spectrum  for masses 
between 2.5 and 3.5\gevcc is performed. The \etac is represented 
by a Breit--Wigner function $(\Gamma/2)^2/((W-m_{\etac})^2+(\Gamma/2)^2)$,
with $W$ the invariant \kskpi mass, convolved with a  Gaussian 
resolution function. The \jpsi peak is fitted with a  Gaussian function. 
The background is represented by an exponential function of $W$, 
$A\exp{(-\lambda W)}$. The free parameters of the fits are the \jpsi mass 
$m_{\jpsi}$, the mass difference $m_{\jpsi}-m_{\etac}$, the \etac width 
$\Gamma_{tot}^{\etac}$, the \jpsi resolution $\sigma_{\jpsi}$, 
the coefficients $A$ and $\lambda$ of the background, 
and the numbers of events in the \etac and \jpsi peaks. The resolution 
$\sigma_{\etac}$ of the \etac peak is constrained to a value 0.8\mevcc 
lower than the \jpsi resolution, as indicated by the Monte Carlo simulation.
The results of the fit are:
$m_{\jpsi}$ = 3093.6 $\pm$ 0.8\mevcc,
$m_{\jpsi}-m_{\etac}$ = 114.4 $\pm$ 1.1\mevcc, 
$\sigma_{\jpsi}$ = 7.6 $\pm$ 0.8\mevcc,
$\Gamma_{tot}^{\etac}$ = 34.3 $\pm$ 2.3\mevcc.
The numbers of \etac and \jpsi events are respectively 2547 $\pm$ 90 and 
358 $\pm$ 33. 

The mass resolution found for the  \jpsi is 0.5 $\pm$ 0.8\mevcc lower than 
the Monte Carlo prediction, but consistent with it. To evaluate the
systematic uncertainty affecting the \etac width, the conditions of the fit are
varied as shown in Table 1. When $\sigma_{\jpsi}$ and $\sigma_{\etac}$ 
are fixed to the values obtained in the Monte Carlo simulation (second row 
of Table 1), the width of the \etac changes by 0.6\mevcc. We take this value 
as an estimate of the systematic uncertainty 
associated with the uncertainty on the \etac resolution. 
The value of $\Gamma_{tot}^{\etac}$ changes by 0.4\mevcc on average when the 
mass range of the fit is varied from 2.4--3.6\gevcc to 2.7--3.3\gevcc. This 
gives an estimate of the systematic uncertainty associated with the choice of 
the mass range of the fit. By varying the event selection parameters, we 
estimate that the systematic uncertainty associated with the event selection 
is 0.5\mevcc. The total systematic uncertainty on the \etac width
 is then 0.9\mevcc. The final value of the \etac width is:
$$\Gamma_{tot}^{\etac} = 34.3 \pm 2.3 (stat) \pm 0.9 (syst)\mevcc.$$

\begin{table}[!h]
\caption{Results of  unbinned maximum likelihood fits to the \etac and \jpsi 
mass spectra. The resolutions of the \jpsi and \etac peaks are respectively 
$\sigma_{\jpsi}$ and $\sigma_{\etac}$. The first row presents the nominal fit, 
and the succeeding rows are used for systematic studies of the \etac width. 
"MC"  denotes results of Monte Carlo simulations.}
\begin{center}\begin{tabular} {|c|c|c|c|} 
\hline
mass range & $\Gamma_{tot}^{\etac}$ & $\sigma_{\jpsi}$ & $\sigma_{\etac}$ \\ 
 \mevcc & \mevcc               & \mevcc         & \mevcc  \\ \hline\hline
2.5--3.5    & 34.3 $\pm$ 2.3    & 7.6 $\pm$ 0.8  & $\sigma_{\jpsi}$-0.8 \\
2.5--3.5    & 33.7 $\pm$ 2.0    & 8.1 (MC)       & 7.3 (MC) \\ \hline
2.4--3.6    & 33.7 $\pm$ 2.3    & 7.6 $\pm$ 0.8  & $\sigma_{\jpsi}$-0.8 \\
2.6--3.4    & 34.4 $\pm$ 2.3    & 7.7 $\pm$ 0.9  & $\sigma_{\jpsi}$-0.8 \\ 
2.7--3.3    & 34.7 $\pm$ 2.4    & 7.7 $\pm$ 0.8  & $\sigma_{\jpsi}$-0.8 \\ \hline
\end{tabular} \end{center}
\end{table}

\begin{figure}[!htb]
\centerline{ \psfig{file=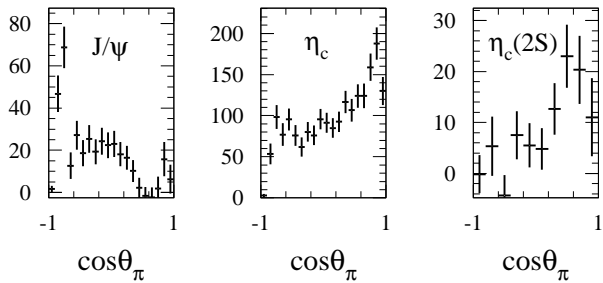,width=8.7cm} }
\caption{Angular distributions of pions from the decays of \jpsi, \etac and
$\eta_c(2S)$, in the laboratory frame ($\Theta_{\pi}$ is the pion polar angle). 
The backgrounds determined from sidebands have been subtracted.}
\label{angpi}
\end{figure}

The \etac mass is 2982.5 $\pm$ 1.1 (stat) \mevcc, obtained by subtracting 
114.4\mevcc from the current world average value of the \jpsi mass~\cite{pdg}.
The \etac and \jpsi masses are unchanged by the alternative fits listed 
in Table 1. We estimate that the systematic uncertainty on
$m_{\jpsi}-m_{\etac}$, associated with the event selection, is 0.8\mevcc.
After correction for the $-1.1\mevcc$ shift seen in simulation, as 
mentioned above, the \jpsi mass is still shifted by an additional 
$-2.2\mevcc$ relative to the well established world average value~\cite{pdg}. 
Because \jpsi events and  \etac events populate different regions of detector 
acceptance, as illustrated in Fig.~\ref{angpi} for final-state pions, a shift 
that applies to the \jpsi may not entirely apply to the \etac due to possible  
imperfections in the detector modelling. When one selects \etac\ events
with decay particles going backward, as is the case for the \jpsi, the \etac\
peak shifts by  0.5\mevcc, which we take as a contribution to the systematic
uncertainty. The final value of the \etac mass is then:
$$m_{\etac} = 2982.5 \pm 1.1 (stat) \pm 0.9 (syst)\mevcc.$$

\begin{figure}[!htb]
\centerline{ \psfig{file=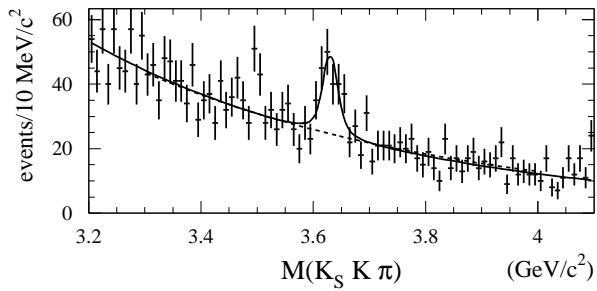,width=8.cm} }
\caption{ The  \kskpi mass spectrum with event selection optimized for 
the $\etac(2S)$ as described in the text. The solid curve is the fit with the 
$\etac(2S)$ resonance shape being represented by a Breit--Wigner 
function convolved with a  Gaussian resolution function. 
The dashed curve shows the background component of this fit.}
\label{etac2s}
\end{figure}

The peak at 3.63\gevcc in the $\KS K \pi$ mass spectrum (Fig.~\ref{fit})
may be  the expected $\etac(2S)$ state. In order to optimize the significance 
of the signal, a new event selection is performed that maximizes the ratio 
$S/\sqrt{B}$. This is appropriate in place of $S/\sqrt{S+B}$ because we need
to establish the significance of the peak without bias from assumptions about
how much signal to expect, and in any case the branching fraction and 
$\gamma \gamma$ width needed for such a prediction are unknown. For $S$ we 
take the signal as generated  from Monte Carlo simulation and $B$ 
is the background estimated from the average of the $\etac(2S)$ sidebands 
3.30--3.48\gevcc and 3.78--3.96\gevcc of the data. Events are required to 
have the total energy deposited by neutral particles less than 0.25\gev and 
$\cos{\theta(\KS)} \geq 0.995$. The other optimized cuts are the same as for
the \etac. The resulting mass spectrum is shown in Fig.~\ref{etac2s}. 

The mass resolution determined from the Monte Carlo simulation is 9.2\mevcc 
and the reconstructed mass is 0.4\mevcc lower than the generated mass. 
Since the resolution for the  \jpsi was found to be 0.5 $\pm$ 0.8\mevcc 
lower in the data than in the Monte Carlo simulation, we assume that the 
resolution for $\etac(2S)$ is also 0.5\mevcc lower in the data, with an 
uncertainty of 0.8\mevcc. The \kskpi 
mass spectrum  is then fitted between 3.3\gevcc and 4.0\gevcc, the $\etac(2S)$ 
resonance shape being represented by a Breit--Wigner function convolved with 
a  Gaussian resolution function with standard deviation 8.7\mevcc. 
The background is fitted with an exponential shape. The fit results in
112 $\pm$ 24 events in the $\etac(2S)$ peak. The significance of this 
signal is characterized by the quantity $\sqrt{2 \times
\log{\cal L}_{max}/{\cal L}_0}$ = 4.9, where ${\cal L}_{max}$ and ${\cal L}_0$ 
are respectively the likelihoods for the fits with and without the 
$\etac(2S)$ peak. 

The $m_{\etac(2S)}-m_{\jpsi}$ mass difference is found to be 534.6 $\pm$ 3.4 
(stat)\mevcc. Taking into account the shifts from generated to reconstructed  
masses of $-1.1\mevcc$ for the \jpsi and $-0.4\mevcc$ for the $\etac(2S)$, as 
found in the Monte Carlo simulation, this mass difference becomes 533.9\mevcc. 
The $\etac(2S)$ mass is then 
$m_{\etac(2S)}$ = $m_{\jpsi}$ + 533.9 = 3630.8 $\pm$ 3.4 (stat)\mevcc.
The measured total width is 17.0 $\pm$ 8.3 (stat)\mevcc. 
The resolution uncertainty of 0.8\mevcc results in a systematic uncertainty
of 0.1\mevcc on the $\etac(2S)$ mass and 2.0\mevcc on its total width. 
When the mass range for the fit is varied to 3.2--4.1 or 3.4--3.9\gevcc, 
the $\etac(2S)$ mass varies by 0.2\mevcc whereas its width varies by 1.2\mevcc 
on average. The $0.5\mevcc$ uncertainty on the 
$-2.2\mevcc$ shift observed for the measured \jpsi mass 
relative to the world average value is taken as a systematic uncertainty on 
the $\etac(2S)$ mass. Based on the upper limit for the branching fraction
$\psi(2S) \to K^+ K^- \pi^0$ \cite{pdg}, we estimate that $\psi(2S)$ (with
a mass of 3.686 \gevcc \cite{pdg}) could contribute up to 5 events to the 
spectrum of Fig.~3. Allowing for this reduces the $\etac(2S)$ width 
by 0.7\mev, which we take as a systematic uncertainty, whereas the $\etac(2S)$ 
mass varies by about 0.1\mevcc. The systematic uncertainties associated with 
the event selection are taken to be the same as for the \etac , 0.8 \mevcc 
for the $\etac(2S)$ mass and 0.5\mevcc for its total width.
Adding all systematic uncertainties in quadrature, the final results are:
$$m_{\etac(2S)} \ = 3630.8 \pm 3.4 (stat) \pm 1.0 (syst)\mevcc$$
$$\Gamma^{\etac(2S)}_{tot} \ = 17.0 \pm 8.3 (stat) \pm 2.5 (syst)\mevcc.$$

\begin{figure}[!htb]
\centerline{ \psfig{file=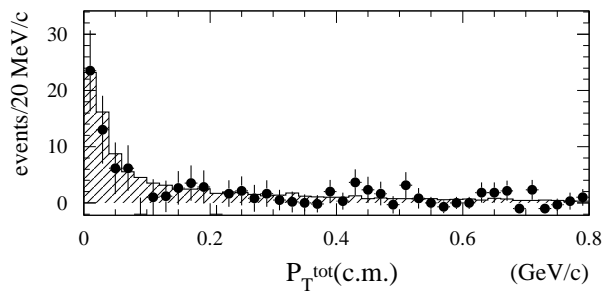,width=8.cm} }
\caption{Total transverse momentum 
in the center-of-mass. The hatched solid 
line is the result of the two-photon Monte Carlo simulation for the $\etac(2S)$
state, normalized to the data. The data are events in the 3.60--3.66\gevcc 
mass region; the background determined from mass sidebands 3.30--3.48\gevcc and 
3.78--3.96\gevcc has been subtracted.}
\label{pttotcm}
\end{figure}

While we have not measured the quantum numbers of the state at 3630.8\mevcc, 
demonstrating that it is formed from the fusion of two quasi-real photons 
would at least restrict the possibilities. Such a process can occur only if 
C=+, and J$^P$ = 0$^-$ (0$^+$ is excluded by the final state), 
2$^\pm$, 3$^+$, 4$^\pm$, ....  Other combinations would be possible if 
production were via an ISR process, or if at least one of the two photons in 
two-photon fusion were highly virtual. However ISR is excluded as the source, 
because the decay products of this state have angular distributions 
concentrated in the forward hemisphere, like the \etac, in contrast to the 
\jpsi for which the decay products peak in the backward direction. This is 
illustrated in Fig.~\ref{angpi}. Moreover the distribution of the total 
transverse momentum (Fig.~\ref{pttotcm}) is peaked at 0, characteristic of 
quasi-real photons, and this excludes spin-one production.
Thus the evidence supports the state having quantum numbers 
$J^{PC}=0^{-+}$ or $J\ge 2$. But $J\ge 2$ is disfavored for a charmonium state 
of such low mass, which suggests that the state has the quantum numbers of the 
$\etac(2S)$.

In summary, we have measured the mass difference between the \jpsi and the 
\etac and the total width of the \etac, using 2547 $\pm$ 90 events of 
$\gamma \gamma \to \etac \to \kskpi$ and  358  $\pm$ 33 
$\jpsi \to \kskpi$ events, selected with the \babar\ detector. 

A state which could be the expected $\etac(2S)$ was also observed in the 
\kskpi decay mode, with 112 $\pm$ 24 events, and its mass and total width 
measured. The measured mass is significantly different from the mass of the 
state reported by the Crystal Ball Collaboration~\cite{edw82}, but consistent 
with the measurements of the Belle Collaboration~\cite{cho02,abe02}. We have 
presented evidence that this 
state is produced via the fusion of two quasi-real photons, which suggests 
that its quantum numbers are those of  the $\etac(2S)$.
The deduced mass splitting $m_{\psi(2S)}-m_{\etac(2S)}$ = 55.2 $\pm$ 
4.0\mevcc is consistent with theoretical expectations.

We are grateful for the excellent luminosity and machine conditions
provided by our \pep2\ colleagues, 
and for the substantial dedicated effort from
the computing organizations that support \babar .
The collaborating institutions wish to thank 
SLAC for its support and kind hospitality. 
This work is supported by
DOE
and NSF (USA),
NSERC (Canada),
IHEP (China),
CEA and
CNRS-IN2P3
(France),
BMBF and DFG
(Germany),
INFN (Italy),
FOM (The Netherlands),
NFR (Norway),
MIST (Russia), and
PPARC (United Kingdom). 
Individuals have received support from the 
A.~P.~Sloan Foundation, 
Research Corporation,
and Alexander von Humboldt Foundation.

\end{document}